\documentclass[aps,twocolumn]{revtex4-1}
\usepackage{graphicx}
\usepackage{wrapfig}
\usepackage{natbib}
\usepackage[toc,page]{appendix}
\usepackage{upgreek}
\usepackage{bm}
\usepackage{amsmath}
\usepackage[colorlinks=true, allcolors=blue]{hyperref}
\usepackage[capitalise]{cleveref}
\usepackage[T1]{fontenc}
\usepackage[utf8]{inputenc}

\newcommand{\micron}{{\upmu \mbox{m}}}
\newcommand{\be}{\begin{equation}}
\newcommand{\ee}{\end{equation}}

\newcommand{\ba}{\begin{eqnarray}}
\newcommand{\ea}{\end{eqnarray}}

\newcommand{\Lf}{Lorentz factor}

\usepackage[dvipsnames]{xcolor}
\newcommand{\rc}[1]{\textcolor{red}{#1}}
\newcommand{\rev}[1]{{\color{black}#1}}

\begin{document}
\title{Complete reflection of nonlinear electromagnetic waves in underdense pair plasmas enabled by dynamically formed Bragg-like structures}

\author{Kavin Tangtartharakul$^1$, Alexey Arefiev$^1$, Maxim Lyutikov$^2$\\
$^1$Department of Mechanical and Aerospace Engineering, University of California San Diego, \\
La Jolla, CA 92093, USA \\
$^2$Department of Physics and Astronomy, Purdue University, \\
 525 Northwestern Avenue,
West Lafayette, IN
47907-2036 }

\begin{abstract} 

\rev{In contrast to relativistically induced transparency in electron--ion plasmas, where nonlinear electromagnetic waves render initially opaque plasmas transparent, we show using kinetic simulations that such waves can instead make initially transparent pair plasmas fully reflective. The difference is mass symmetry, which eliminates charge-separation electric fields. As the wave compresses the pair plasma, weak reflection seeds density spikes that form a moving Bragg-like grating. Enhanced reflection enables a transition to a regime where the plasma--vacuum interface sustains complete reflection.}
\end{abstract}



\maketitle 

Fast Radio Bursts (FRBs) are extremely bright, brief flashes of radio waves observed across cosmological distances~\citep{Lorimer.Science.2007.FRB,Petroff.AAPR.2022.FRB,Lorimer.ASS.2024.Discovery}, whose origin remains a subject of active research. A key aspect of FRBs is understanding how intense electromagnetic waves strong enough to accelerate particles to relativistic speeds propagate through the surrounding plasma~\cite{Qu.RAS.2022.Transparency}. In many scenarios, the plasma consists of electrons and positrons rather than electrons and ions~\cite{Brambilla.TAJ.2018.Pulsar}, raising questions about the transparency of pair plasmas to high-amplitude waves~\cite{2021ApJ...922L...7B,2024MNRAS.529.2180L,Huang.PPCF.2021.Pair,Sobacchi.PhysRevResearch.2024}.

There is a parallel between FRBs and laboratory systems driven by high-intensity lasers. Recent advances have enabled multi-petawatt laser facilities to produce pulses strong enough to drive relativistic laser–plasma interactions~\cite{Danson.HPLSE.2019.Worldwide,Cheriaux.AIP.2017.APOLLON,Yoon.Optica.2019.CoReLS,Doria.JI.2020.ELINP,Borneis.HPLSE.2021.ELIbeamlines}. This allows experiments to access regimes once thought exclusive to extreme astrophysical settings~\cite{Stark.PRL.2016.Emission,Chen.PoP.2023.Perspectives,Raymond.PRE.2018.Reconnection,Zhang.PoP.2020.Supercritical,Gonoskov.2022.RMP.Motion}. However, laboratory plasmas consist of electrons and ions, whereas many FRB scenarios involve electrons and positrons. We show that this difference fundamentally alters wave propagation by changing plasma transparency, so even basic aspects of FRB transmission cannot be reliably inferred from electron-ion systems.




In an electron–ion plasma, wave propagation changes once the electromagnetic wave amplitude becomes large. There are two distinct regimes, determined by the normalized amplitude $a_0 \equiv |e| E_0 / m_e c \omega$, where $E_0$ is the wave's electric field amplitude, $\omega$ is its frequency, $c$ is the speed of light, and $e$ and $m_e$ are the electron charge and mass. In the linear regime ($a_0 \ll 1$), the plasma is transparent if the electron density $n_e$ is below the critical density, $n_{cr} = m_e \omega^2 / 4\pi e^2$, and opaque otherwise. In the nonlinear regime ($a_0 \gg 1$), the wave drives relativistic electron motion, effectively increasing the electron mass by a factor $\gamma \sim a_0$. This raises the cutoff density by the same factor -- a phenomenon known as relativistically induced transparency (RIT)~\cite{gibbon_WorldScientific_2005, akhiezer}. \rev{RIT has been observed in experiments~\cite{palaniyappan_NaturePhysics_2012, Fernandez_PoP_2017} and is supported by theory and simulations~\cite{Cattani.PRE.2000, stark.prl.2015, Siminos.NJP.2017}. In a pair plasma, the cutoff density is reduced by a factor of two at $a_0 \ll 1$ due to the additional positron contribution to the plasma response. One might then also expect a wave with $a_0 \gg 1$ to cause RIT in a pair plasma, with the density threshold simply reduced by a factor of two. We find that the opposite takes place: an initially transparent pair plasma becomes opaque at $a_0 \gg 1$. The reason is the mass symmetry, which eliminates charge-separation electric fields that play a critical role in electron-ion plasmas.}

\begin{figure}
 \includegraphics[width=.85\linewidth]{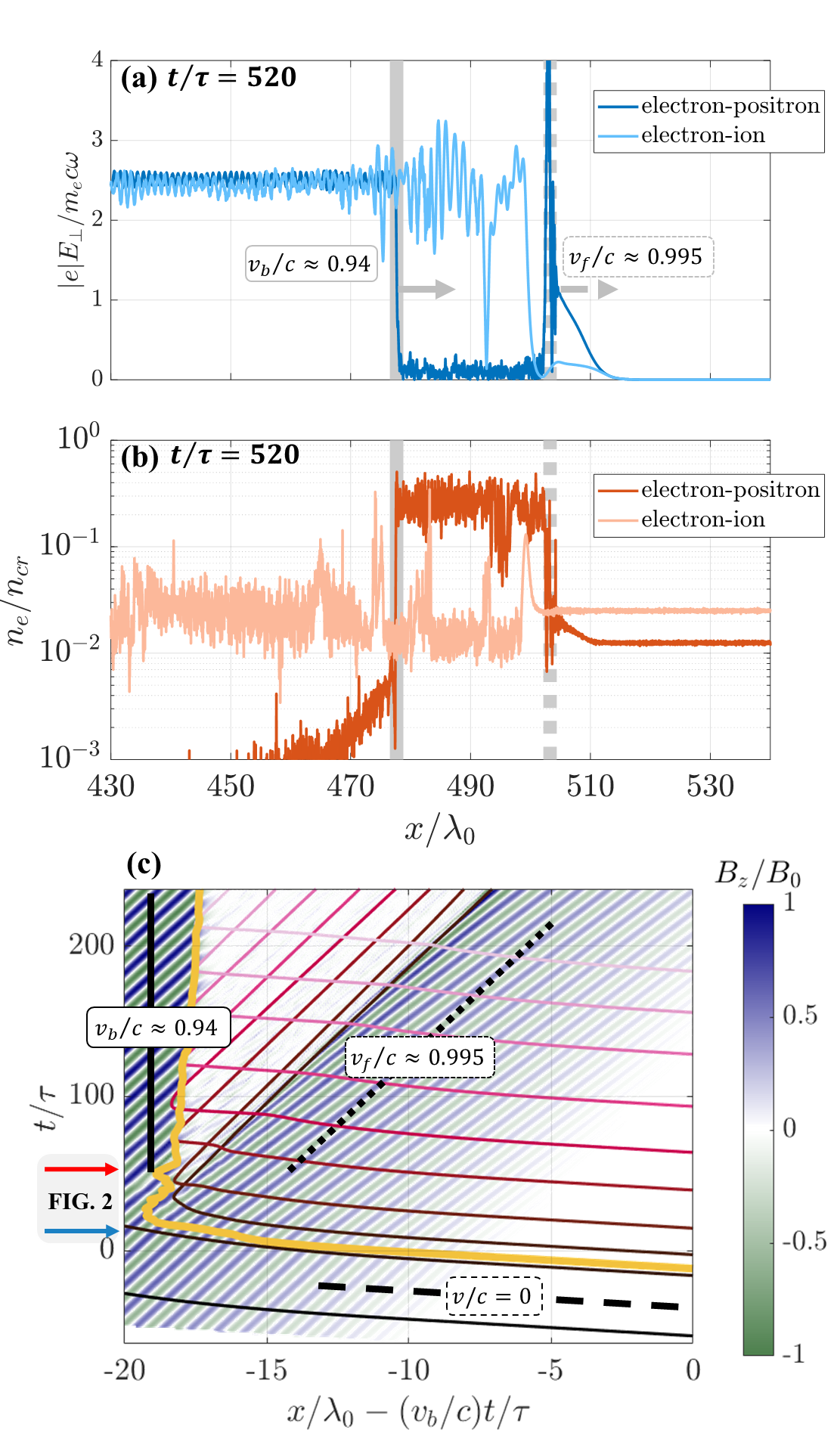}
\caption{1D PIC simulations of a circularly polarized wave incident on an electron-ion and an electron-positron (pair) plasma. (a,b) Transverse electric field amplitude $E_{\perp}$ and electron density $n_e$. A reflection front (solid gray line) forms in the pair plasma, supported by a dense reflective plug. (c) Space-time map of $B_z/B_0$ and sample electron trajectories in the pair plasma, where $B_0$ is the peak magnetic-field amplitude of the incident wave. The yellow curve marks the leftmost position where the density first reaches 50\% of its maximum. Trajectories are color-coded by initial position: lighter colors = initially farther from the boundary.} \label{fig:circ_plot}
\end{figure}

\rev{\Cref{fig:circ_plot} shows a qualitative difference in plasma response at $a_0 = 2.5$. The results come from 1D particle-in-cell (PIC) simulations using the parameters listed in \cref{sec:app: EPOCH}. The initial electron density is $n_e = 0.025 n_{cr}$ in the electron–proton case and half that in the pair plasma. In the electron–ion plasma, \cref{fig:circ_plot}(a) shows that the pulse propagates into the plasma, with the density profile shown in \cref{fig:circ_plot}(b). In contrast, in the pair plasma, the wave is completely reflected. The reflection takes place at a moving boundary, visible as a density jump in \cref{fig:circ_plot}(b) and marked by a solid gray line. It occurs even though $n_e \ll a_0 n_{cr}/2$, underscoring that the observed effect is the opposite of RIT in an electron--ion plasma.}

\rev{Our key finding is that the complete reflection by an underdense pair plasma with $n_e < n_{cr}/2$ results from the formation of a relativistically moving Bragg-like grating enabled by the mass symmetry of electrons and positrons. At $a_0 > 1$, the wave sweeps up the plasma, compressing and accelerating it to relativistic velocity. In this flow, spikes form as an instability is triggered by a weak reflected wave, which produces a pinching force together with the incident wave. Even as the spike density remains below $n_{cr}$, the moving periodic structure acts as a photonic crystal that strongly reflects the wave. Because the spikes travel with the relativistic plasma flow, only a few are needed to sustain reflection, unlike in a static grating. The enhanced reflection increases momentum transfer, driving the boundary to higher velocity and enabling a transition to a regime of complete reflection.}


\rev{A key aspect of a wave with $a_0>1$ interacting with a pair plasma is that it compresses the plasma without much pushback. At $a_0>1$, electrons and positrons both become relativistic and, as a result, experience a strong force from the wave’s magnetic field that pushes them forward. Because of the mass symmetry, no charge separation occurs during this motion. By contrast, in an electron--ion plasma the ions do not respond directly to the wave, so only the electrons experience the compression. This generates a charge-separation electric field that counteracts electron compression and also prevents the plasma sweep-up that we observe in a pair plasma.}

\rev{To quantify the impact of the plasma compression, we analyze a simplified scenario in which a circularly polarized wave with \( a_0 \gg 1 \) propagates through a low-density, cold, uniform pair plasma.} The low-density assumption allows us to treat the wave’s phase velocity \( v_{ph} \) as equal to \( c \) when calculating the particle dynamics. This is justified if the actual phase velocity satisfies \( (v_{ph} - c)/c \ll 2/a_0^2 \)~\cite{robinson.pop.2015}. In this regime, the particle dynamics is well known~\cite{gibbon_WorldScientific_2005}. Using this solution, we apply a Lagrangian formulation to calculate \( \partial x / \partial x_0 \), where \( x_0 \) is the initial position, and then determine the local electron and positron densities. This standard approach (details in \cref{sec:app: Pond dens enhance}) yields $n = \gamma n_0$, with $\gamma = 1 + a_0^2/2$. Thus, the density of each species increases by $\gamma$.

We can now examine how the density compression modifies the wave itself.  After expressing the transverse current \(j_{y,z}\) through the single-particle solutions for \(p_{y,z}\)~\cite{gibbon_WorldScientific_2005} and substituting it into Maxwell’s equations, written in the Coulomb gauge where \( \bm{E} = - c^{-1} \partial \bm{A}/\partial t \) and \( \bm{B} = \nabla \times \bm{A} \), we obtain the wave equations for the two components of the normalized vector potential, $\bm{a} = |e|\bm{A}/ m_e c^2 $,
\begin{equation}
  \label{eq:ay_eom}
  \frac{1}{c^{2}}\frac{\partial^{2} a_{y,z}}{\partial t^{2}}
  -\frac{\partial^{2} a_{y,z}}{\partial x^{2}}
  = -\,\frac{2\,\omega_{p}^{2}}{c^{2}\,\gamma}\;a_{y,z},
\end{equation}
where \(\omega_{p}^{2}=4\pi n e^{2}/m_{e}\). The analysis is particularly simple for the circularly polarized wave, because there are no density or $\gamma$ modulations and the equation becomes effectively linear. For a wave with amplitude $a_0$, frequency $\omega$, and wavenumber $k$, we find that $\omega^2 - k^2 c^2 = 2 \omega_{p}^2 / \gamma$. Substituting the compressed density $n = \gamma n_0$ and using the definition of $n_{cr}$, we obtain
\begin{equation} \label{eq: normal}
    \omega^2 = k^2 c^2 + 2\omega_{p0}^2 = k^2 c^2 + (2 n_0/n_{cr}) \omega^2 , 
\end{equation}
\rev{where \(\omega_{p0}^{2}=4\pi n_0 e^{2}/m_{e}\). Thus, the compression exactly offsets the effect of relativistic transparency. For our parameters, $(v_{ph} - c)/c \ll 2/a_0^2$, confirming self-consistency.}

\rev{In an electron--ion plasma, by contrast, electron compression does not cancel relativistic transparency, but instead shifts the threshold for its onset~\cite{Cattani.PRE.2000}. Ion motion in the charge-separation electric field can further modify the plasma response~\cite{Siminos.NJP.2017}. However, the main trend remains the same: an initially opaque electron--ion plasma becomes transparent at sufficiently large wave amplitude. Our result for the pair plasma is the opposite.}


\begin{figure} 
  \centering\includegraphics[width=.95\linewidth]{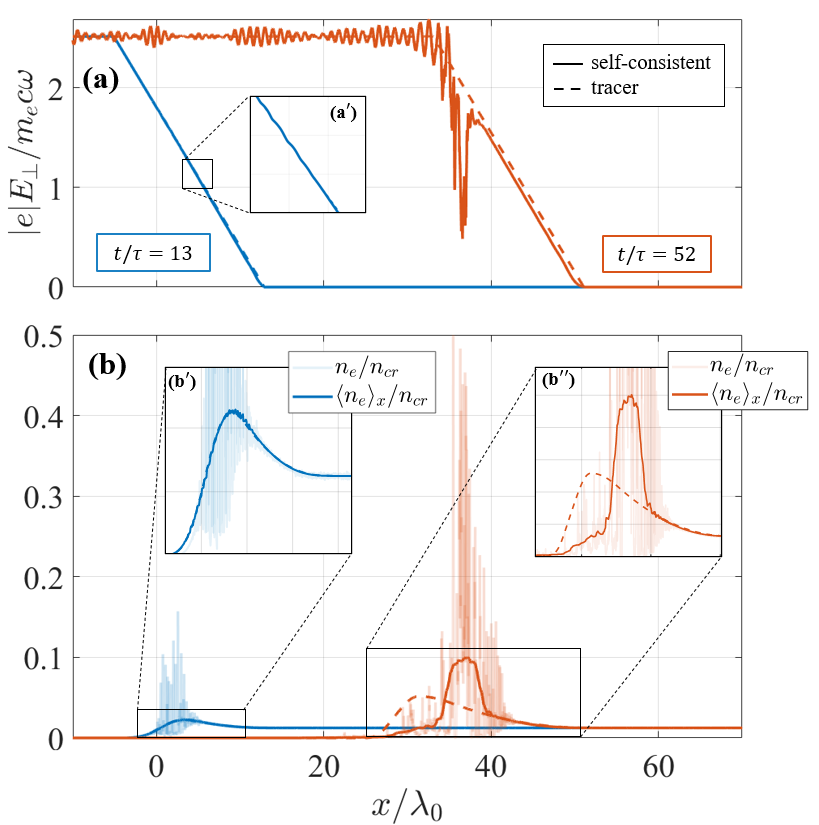}
  \caption{Comparison of the pair plasma simulation from \cref{fig:circ_plot} with a tracer-particle simulation (dashed) where the wave is unaffected by the plasma. (a) Transverse electric field. (b) Electron density: faint curves are the raw $n_e$, while thick curves are the averaged density $\langle n_e \rangle_x$.}
\label{fig: breakdown}
\end{figure}



\Cref{fig: breakdown} shows the wave amplitude and electron density at \( t = 13\tau \), where \(\tau\) is the laser period, comparing the self-consistent simulation with a tracer-particle simulation. In the latter, the particles respond to the wave but do not affect it. At \( t = 13\tau \), the wave is compressing the plasma, but the raw density $n_e$ in panel (b) differs markedly from the analytical prediction. To gain insight, we computed the averaged density \(\langle n_e \rangle_x\). Here and in what follows, $\langle \cdot \rangle_x$ denotes a spatial sliding-window average over four vacuum wavelengths $\lambda_0$ used to filter out small-scale oscillations. This average coincides with the tracer-particle simulation (hence the tracer curve is not visible), which itself follows the analytical solution. We therefore conclude that we have sharp density peaks developing on top of the compressed plasma.




The spikes form because an instability is triggered by even a very weak reflected wave inside the plasma. To illustrate the mechanism, we consider two counter-propagating circularly polarized waves of equal amplitude, frequency $\omega$, and wavenumber $k$, with the same rotation, in the nonrelativistic limit. A single wave produces no net longitudinal force, but the pair does. The longitudinal momentum of a particle obeys \( dp_x/dt = q [\bm{v}_\perp \times \bm{B}_\perp] / c\), where \(\bm{v}_\perp\) is the transverse velocity and \(\bm{B}_\perp\) is the total magnetic field. For the two-wave field, this gives $dp_x/dt = -2 a_0^2 m_e c \omega \sin(2kx)$. Thus, particles are pinched at spacing $\lambda/2 \equiv \pi/k$~\cite{Palastro_2015, Lyutikov_2021}.



\begin{figure}
 \centering
 \includegraphics[width=.99\linewidth]{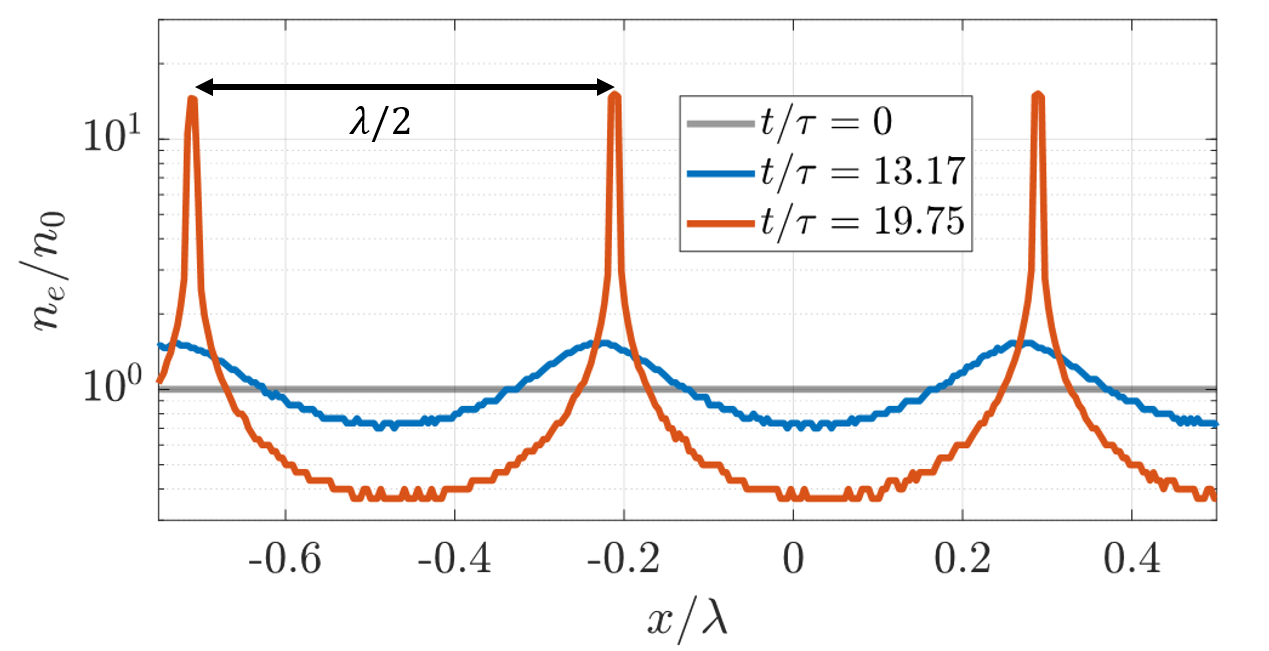}
 \caption{Electron density (normalized to the initial value $n_0$) in test-particle simulations with two counter-propagating circularly polarized waves (forward $a_0=0.1$, backward $a_0=10^{-3}$). A pinching force forms spikes with spacing $\lambda/2$.} \label{fig: scanx}
\end{figure}

This pinching produces periodic density spikes. We demonstrate this with a test-electron simulation using two prescribed circularly polarized waves, with the counter-propagating wave two orders of magnitude weaker than the primary. The electron-density in \cref{fig: scanx} shows that, despite the very weak counter-propagating component, spikes grow rapidly — reaching an order of magnitude above the initial density within 20 periods.

The described scenario plays out in the plasma rest frame moving with $\beta_{\parallel} \approx a_0^2/(2 + a_0^2)$. To find the spacing in the lab frame, note that for a strongly underdense plasma the rest-frame frequency (primes denote this frame) is
\begin{equation} \label{omega'}
    \omega' \approx \omega (1 - \beta_{\parallel})^{1/2}/(1 + \beta_{\parallel})^{1/2}.
\end{equation}
In this frame, the spikes are static with spacing $\lambda'/2$, where $\lambda' = 2\pi c/\omega'$. Lorentz contraction along the motion gives the lab-frame spacing $\Delta x = \lambda'/2 \gamma_{\parallel} = \lambda_0 (1+\beta_\parallel)/2$, with $\gamma_{\parallel} = (1 - \beta_{\parallel}^2)^{-1/2}$. Thus, $\Delta x$ increases with $\beta_{\parallel}$ and exceeds $\lambda_0/2$, consistent with the simulations.

The spikes act as a photonic crystal, reflecting the wave. \rev{There are parallels with plasma photonic structures studied in electron--ion plasmas~\cite{lehmann.prl.2016,vieux.commphys.2023}. However, here the structure is not externally prepared by interfering driver pulses, but forms self-consistently from a single incident wave. Moreover, the structure moves with the plasma, so we analyze it in the plasma rest frame and model the spike–trough pattern as a periodic array of cells, each composed of two uniform slabs: a high-density “spike” slab and a low-density “trough” slab, with particle number per cell conserved.} The cell width is half the local wavelength for the density before spike formation. The dispersion relation for the plasma without spikes is $(\omega')^2 = (k')^2 c^2 + 2\omega_{p0}^2$, so the dielectric constant is $\varepsilon' = 1 - N'$ with $N' \equiv (2n_0/n_{cr}) (1 + \beta_{\parallel}) /(1-\beta_{\parallel})$. The pinching acts as a modulation of $N'$, so the standard transfer-matrix analysis for reflection applies (given in \cref{sec:grating} for completeness).

As an example, assume that half the particles in each cell are compressed fivefold, while the rest are decompressed. For $a_0 = 2.5$ and $n_0 = 0.0125 n_{cr}$ ($\beta_\parallel \approx 0.76$), the resulting modulation of $\varepsilon'$ reduces the field amplitude across a cell to $e^{-0.15} \approx 0.86$ of its initial value. Only five spikes are then needed to halve the amplitude, producing strong reflection. In the lab frame, the density in the dense slabs (the spikes) remains well below critical: $n = 5 \gamma n_0 \approx 0.26 n_{cr}$. Strong reflection from only a few spikes is a consequence of the rest-frame frequency lowering given by \cref{omega'}.

\begin{figure}
 \centering
\includegraphics[width=\linewidth]{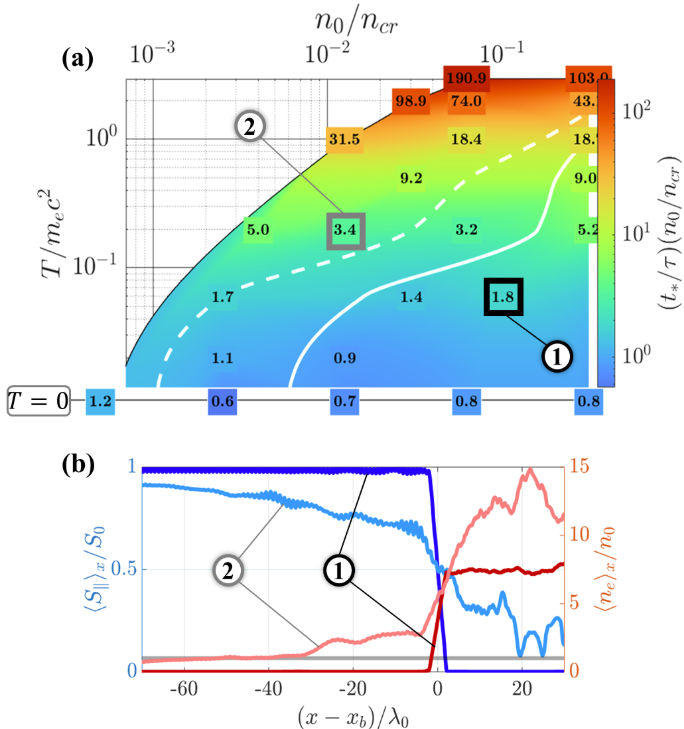}
\caption{(a) $T$ vs $n_0$ scan showing the fraction of swept-up plasma leaking through the density plug (contours: 5\% solid; 20\% dashed) and the time to reach a 90\% reduction of the Poynting flux $\langle S_{\parallel}\rangle_x$ (color). (b) Snapshots of $\langle S_{\parallel} \rangle_x$ and $\langle n_e \rangle_x$ near the density plug located at $x_b$ for cases (1) $n_0 = 0.1 n_{cr}$, $T=0.06 m_e c^2$ and (2) $n_0 = 0.0125 n_{cr}$, $T=0.2 m_e c^2$. For these snapshots, (1) $t/\tau = 285$, $x_b/\lambda_0 \approx 229$ and (2) $t/\tau = 875$, $x_b/\lambda_0 \approx 601$.} \label{fig: temp_plots}
\end{figure}

As seen in \cref{fig: breakdown}(a), the reflection induced by the spikes becomes evident in the field profile by $t \approx 52 \tau$. The reflection transfers additional momentum from the wave to the plasma -- a process absent without reflection. The extra momentum manifests as further plasma compression and acceleration. In \cref{fig: breakdown}(b), the red density profile shows this effect: the thick solid curve for the averaged density $\langle n_e \rangle_x$ lies well above $n_e$ from the tracer-particle simulation (dashed curve). The density peak is also displaced forward, a result of the stronger push imparted by the reflected field.

The additional momentum transfer associated with the reflection alters the particle dynamics, as seen in \cref{fig:circ_plot}(c) for $t \gtrsim 50 \tau$. While the plasma is initially compressed without particle overtaking, once reflection sets in, particles are turned back at the boundary (solid yellow curve). Two concurrent effects then reduce transparency: the boundary accelerates, which lowers the wave frequency in its rest frame, and the turned-back particles form a counterstream, increasing the plasma density. 

We quantify these effects by considering the regime that sets in after the initially compressed layer has been turned back and cleared from the boundary region. The boundary, moving at $\beta_b$, then encounters stationary plasma with the original density $n_0$ and deflects particles forward. In the boundary rest frame (subscript $b$ denotes this frame), the plasma consists of two counter-streaming flows of equal density $n_b = \gamma_b n_0$ moving with $\pm \beta_b$, where $\gamma_b = (1-\beta_b^2)^{-1/2}$. For simplicity, we assume $\gamma_b \gtrsim a_0$, so the dielectric constant for this setup is just $\varepsilon_b = 1 - N_b$, with $N_b \equiv 4n_b/\gamma_b n_{cr,b}$ \cite{arefiev.pop.2020}. After taking into account that $n_{cr,b} = m_e \omega_b^2 / 4\pi e^2$ and $\omega_b \approx \omega (1 - \beta_b)^{1/2}/(1 + \beta_b)^{1/2}$,  we express $N_b$ in terms of $n_0$ and $n_{cr}$ as $N_b \equiv (4n_0/n_{cr})(1 + \beta_b)/(1 - \beta_b)$. The expression for $N_b$ explicitly quantifies the reduction in transparency discussed in the previous paragraph.

Using a standard momentum-balance analysis \cite{Robinson.PPCF.2009}, we can determine when the reduced transparency produces a reflection front. Balancing the electromagnetic and plasma momentum fluxes in the boundary rest frame (see Appendix~\ref{sec:app:rel hole boring} for intermediate steps) gives $\beta_b = 1/(1+\sqrt{\psi})$, with $\psi \equiv 2 n_0/a_0^2 n_{cr}$. The front is reflective if $N_b > 1$. Substituting $\beta_b$ into $N_b$ and taking the underdense limit ($\psi \ll 1$), we obtain \begin{equation} \label{threshold}
    n_0/n_{cr} > 1/32 a_0^2,
\end{equation}
which we will refer to as the threshold condition.




The pair-plasma simulation in \cref{fig:circ_plot} is captured by this model. For the simulated parameters, $\psi \approx 4 \times 10^{-3}$, giving $\beta_b \approx 0.94$ ($\gamma_b \approx 3 \gtrsim a_0$) and $N_b \approx 1.6$. The simulated regime satisfies the threshold condition \cref{threshold}, so a reflection front forms, shown by the yellow curve in \cref{fig:circ_plot}(c). Since the leading edge of the wave entered the plasma before the reflection front formed, a region with embedded fields remains. However, the field-free region (``blank wedge'') expands with time as the reflected particles outrun the reflection front.

\rev{The compression is a key element and, in the absence of a charge-separation electric field, thermal motion is the main mechanism that can impede it in a pair plasma. To quantify its impact, we performed a temperature scan for a range of densities $n_0$ using a pair plasma initialized with a Maxwell--Jüttner distribution at temperature $T$. Increasing $T$ shifts the full-reflection threshold \cref{threshold} to higher $n_0$, while at $T = 0$ the observed threshold matches the prediction. Above threshold, the morphology is unchanged across the temperatures we examined: a dense plug forms, fully reflecting the wave. Case (1) in \cref{fig: temp_plots}(b), with $n_0 = 0.1 n_{cr}$ and $T=0.06 m_e c^2$, illustrates this: the density $\langle n_e \rangle_x$ and the Poynting flux $\langle S_\parallel \rangle_x$ profiles resemble the $T=0$ case. The profiles are plotted relative to the plug location $x_b$, defined as the leftmost position where $\langle S_{\parallel} \rangle_x /S_0 = 0.5$, with $S_0$ being the incident Poynting flux.

At densities below the threshold, the interaction still produces a plug rather than rendering the plasma fully transparent. The plug, however, is leaky, with a fraction of the swept-up plasma passing through. Case (2) in \cref{fig: temp_plots}(b), with $n_0 = 0.0125 n_{cr}$ and $T=0.2 m_e c^2$, illustrates this: the density in the plug increases by a factor of 15 while the Poynting flux drops to $0.1 S_0$, with $\sim 20 \%$ leakage. This mirrors the cold, below-threshold case at $n_0 = 2.5 \times 10^{-3} n_{cr}$ (not shown). In \cref{fig: temp_plots}(a), contours mark 5\% and 20\% leakage across the scan at times after the plug is well established. Strong modulations of $\langle S_\parallel \rangle_x$ are observed throughout the scan, with the color in \cref{fig: temp_plots}(a) showing the time to reach $0.1 S_0$. To the right of the dashed $20\%$ leakage contour, the time to reach $0.1 S_0$ coincides with the plug-formation time. In contrast to electron--ion plasmas, where relativistic thermal motion promotes transparency~\cite{stark.prl.2015, arefiev.pop.2020}, the key role of thermal motion here is to weaken the compression.}

\rev{In summary, high-$a_0$ waves in pair plasmas produce the opposite response to relativistically induced transparency in electron--ion plasmas. Instead of rendering an initially opaque plasma transparent, they can make an initially transparent pair plasma fully reflective. The key is the mass symmetry of electrons and positrons, which eliminates the charge-separation electric field that counteracts compression in an electron--ion plasma. Relativistic transparency is prevented because the wave organizes a dense plug at its front edge, aided by a self-consistently formed, relativistically moving Bragg-like grating of density spikes. While we used circular polarization for simplicity, additional simulations with linear polarization show the same effect. Our 1D analysis is well suited to astrophysical environments with locally planar wavefronts. Thermal motion raises the full-reflection threshold, yet the corresponding density can remain well below conventional relativistic-transparency scaling, even at relativistic temperatures.}

We thank Yong Cheng and Amitava Bhattacharjee for fruitful discussions. K.T. and A.A. were supported by the National Science Foundation (PHY-2512067, PHY-2206777) and by the Air Force Office of Scientific Research (FA9550-24-1-0053). This research was supported in part by grant NSF PHY-2309135 to the Kavli Institute for Theoretical Physics (KITP). This work used Stampede3 at TACC through allocation TG-PHY250187 from the Advanced Cyber Infrastructure Coordination Ecosystem: Services \& Support (ACCESS) program, which is supported by National Science Foundation grants \#2138259, \#2138286, \#2138307, \#2137603, and \#2138296. Simulations were performed using EPOCH, which was developed as part of the UK EPSRC-funded projects no. EP/G054950/1, EP/G056803/1, EP/G055165/1 and EP/ M022463/1.


\bibliographystyle{apsrev4-1}
\bibliography{./BibTexShort,Collections} 

\clearpage
\appendix


\section{Parameters used in PIC simulations} \label{sec:app: EPOCH}


Table \ref{table: pic} provides the parameters used in our 1D-3V PIC simulations. All simulations were performed using the fully relativistic PIC code EPOCH \cite{arber.ppcf.2015}. While the physics is governed by normalized quantities, EPOCH requires dimensional quantities for input. For reproducibility, we list in the table the specific dimensional values used in our simulations, which match the normalized parameters discussed in the main text.

\begin{table} [b!]
\begin{tabular}{ | p{4.5cm} | p{4.0cm} |}
\hline
\multicolumn{2}{|c|}{Laser parameters} \\
\hline
Polarization & Circular \\
\hline
Peak normalized  amplitude & $a_0 = $ 2.5\\
\hline
Peak electric field & $E_0 =$8.0$\times 10^{12}$~V/m \\
\hline
Peak magnetic field & $B_0 = $2.7$\times 10^{4}$~T \\
\hline
Vacuum wavelength & $\lambda_0 = 1.0~\micron$\\
\hline
Wave period & $\tau =3.33$ fs\\
\hline
Propagation direction & $+x$ \\
\hline
Temporal field profile & Linear leading-edge ramp up over $18\tau$\\
\hline
\hline
\multicolumn{2}{|c|}{Pair-plasma parameters (baseline simulation)} \\
\hline
Composition & Electrons and positrons \\
\hline
Bulk electron density & $n_0= 0.0125 n_\mathrm{cr}$\\
\hline
Critical density & $n_{cr} = 1.1 \times~10^{21}$ cm$^{-3}$\\
\hline
Density ramp & $\exp[-x^2/(6 \lambda_0)^2]$ for $x<0$ \\
\hline
Spatial resolution & 100 cells/$\lambda_0$ \\
\hline
Macroparticles per cell & 100 for electrons\\
       & 100 for positrons\\
\hline
Simulation box [$x_{min},x_{max}$] & [-60,590] $\micron$ \\
\hline
\end{tabular}
\caption{Parameters used in 1D-3V PIC simulations.}
\label{table: pic}
\end{table}

In the baseline pair-plasma simulation of Fig.~\ref{fig:circ_plot}, the target is a semi-infinite slab consisting of electrons and positrons initialized with zero temperature (\(T = 0\)). The initial electron density follows a semi-Gaussian ramp for \(x < 0\) [see Table~\ref{table: pic}], transitioning to a uniform bulk density \(n_e = n_0\) for \(x \ge 0\). The positron density is equal throughout. The electromagnetic wave propagates in the positive \(x\)-direction, with \(t = 0\) defined as the moment its leading edge reaches \(x = 0\). Its normalized amplitude increases linearly over the first \(18\tau\) (where \(\tau\) is the laser period) before remaining constant at \(a_0\). The simulation employed open boundary conditions for the electromagnetic fields and for the particles.

In the electron–proton plasma simulation shown in Fig. \ref{fig:circ_plot}, we used exactly the same laser parameters as in the baseline pair-plasma simulation and the same boundary conditions (for fields and particles). The initial electron density followed the same spatial profile. The density at each position was doubled, so that the uniform bulk density was $n_0=0.025 n_\mathrm{cr}$. The proton density was equal to the electron density. The number of macroparticles per cell representing protons was the same as the number of macroparticles per cell representing electrons. Both electrons and protons were initialized with zero temperature (\(T = 0\)).

The scan in Fig.~\ref{fig: temp_plots} is obtained by performing PIC simulations across a range of densities and temperatures. In contrast to the baseline simulation, a uniform pair plasma filled the entire simulation window to prevent plasma expansion into the vacuum gap (present in the original setup) prior to the arrival of the wave, which would otherwise reduce the plasma density. The right boundary of the simulation box was set to be reflective for particles. The left boundary was open, but this had little effect, as the wave quickly pushed the particles to the right. We still used open boundary conditions for the fields. In the scan, the density was varied from \(n_0/n_\mathrm{cr} = 5 \times 10^{-4}\) to \(n_0/n_\mathrm{cr} = 0.3125\). The plasma distribution was initialized using a Maxwell--J\"uttner distribution, and the temperature was varied from \(T/mc^2 = 0.06\) to \(T/mc^2 = 3\). For simulations requiring extended run-times, we employed a simulation window moving at the vacuum speed of light in the \(+x\) direction.

\section{Plasma pile-up} \label{sec:app: Pond dens enhance}

We consider a slab of cold, uniform pair plasma with initial electron and positron density $n_0$, irradiated by a normally incident plane electromagnetic wave propagating along the $x$-axis. Our goal is to calculate the density pile-up under the assumption that the plasma density is sufficiently low to neglect the plasma impact on the wave.

Under this assumption, the wave fields are prescribed, allowing the particle dynamics to be determined directly. Plasma electrons and positrons experience an identical longitudinal force from the magnetic field of the wave, so no charge separation develops. As a result, the dynamics reduces to the well-known solution for a free electron or positron irradiated by a plane wave~\cite{gibbon_WorldScientific_2005}. For a linearly polarized wave with normalized vector potential $a_{\perp} (\xi)$, where $\xi = \omega (t - x/c)$ is the phase variable, we have
\begin{equation} \label{eq: single electron}
    \frac{p_{\perp}}{m_e c} = - \frac{q}{|e|} a_{\perp}, \mbox{  } \frac{p_{\parallel}}{m_ec}  = \frac{a_{\perp}^2}{2}, \mbox{  }  \gamma = 1 + \frac{p_{\parallel}}{m_e c}. 
\end{equation}
Here, $p_{\perp}$ and $p_{\parallel}$ are the transverse and longitudinal momenta; $\gamma$ is the relativistic factor; $q$ is the particle charge ($\pm |e|$); $m_e$ is the electron mass; and $c$ is the speed of light.

The position of a particle starting at $x_0$ is given by
\begin{equation}
    x(t) = x_0 + \int_{t_0}^t v_{\parallel} dt',
\end{equation}
where $v_{\parallel} = dx/dt$ is the longitudinal velocity. We recast this expression using the phase of the wave, $\xi = \omega (t - x/c)$. To express $dt$ in terms of $d \xi$, we calculate 
\begin{equation}
  \frac{d \xi}{dt} = \omega \left[ 1 - \frac{v_{\parallel}}{c} \right] = \omega / \gamma,
\end{equation}
where we used $v_{\parallel} = p_{\parallel}/\gamma m_e$ and then expressed $p_{\parallel}$ in terms of $\gamma$ using Eq.~(\ref{eq: single electron}). Substituting $dt = \gamma d\xi / \omega$ and rewriting $v_{\parallel} = p_{\parallel}/\gamma m_e$, gives
\begin{equation}  \label{B5}
    x(t) = x_0 + \int_{\xi_0}^{\xi} \frac{p_{\parallel}}{m_e \omega} d \xi' .
\end{equation}

The derived expression for $x(t)$ can now be used to compute the density. When the particles move without overtaking each other, particle conservation implies that $n \Delta x = n_0 \Delta x_0$, where $\Delta x$ is the length of the spatial interval currently occupied by particles that were initially located within an interval of length $\Delta x_0$. In  the limit $\Delta x_0 \rightarrow 0$, this relation yields the following expression for the particle density:
\begin{equation} \label{eq: n - 1}
  n = n_0 \left[ \partial x/ \partial x_0 \right]^{-1},
\end{equation}
where $x$ is the position of a particle initially located at $x_0$. We find from Eq.~(\ref{B5}) that
\begin{equation} \label{eq: dx/dx0} 
    \frac{ \partial x}{\partial x_0} = 1 + \frac{\partial \xi}{\partial x_0} \frac{p_{\parallel}}{m_e \omega} = 1 - \frac{ \partial x}{\partial x_0} \frac{p_{\parallel}}{m_e \omega}.
\end{equation}
Solving for $\partial x/\partial x_0$ and using the expression for $\gamma$ from Eq.~(\ref{eq: single electron}), we obtain
\begin{equation}  
    \partial x/\partial x_0 = \gamma^{-1},
\end{equation}
which leads to the final result:
\begin{equation} \label{eq:n-2}
  n = \gamma n_0.
\end{equation}
This result shows that the plasma becomes compressed, with the density increasing by a factor equal to the relativistic factor $\gamma$ of the accelerated particles.

Although the derivation above assumes a uniform initial density \( n_0 \), the same approach applies to an arbitrary initial profile \( n_0(x_0) \), provided the particle trajectories do not cross. In that case, the mapping between initial and final positions remains one-to-one, allowing the density evolution to be calculated from the conservation of the number of particles in a phase space element, just as in the uniform case.


\section{Reflection by Bragg grating} \label{sec:grating}

We follow a standard transfer–matrix analysis, considering what we will refer to as a single cell: two plasma layers with dielectric constants $\varepsilon_1$ and $\varepsilon_2$ and thicknesses $l_1$ and $l_2$. The wave frequency is $\omega$. In each homogeneous layer, the wave satisfies
\begin{equation}
k_i^2 = \varepsilon_i \frac{\omega^2}{c^2},
\end{equation}
so the wavenumber is $k_i = (\omega/c)\sqrt{\varepsilon_i}$. Without any loss of generality, we consider a linearly polarized wave with electric field $E_y$ and magnetic field $B_z$. 

Inside layer $i$, the fields are given by
\begin{align}
E_y(x,t) &= E_+ e^{ik_i x - i\omega t} + E_- e^{-ik_i x - i\omega t}, \\
B_z(x,t) &= \frac{k_i c}{\omega}\left[E_+ e^{ik_i x - i\omega t} - E_- e^{-ik_i x - i\omega t}\right],
\end{align}
which satisfy Maxwell’s equations. The fields on the right side of the layer are given in terms of the fields on the left side of the layer by
\begin{equation}
\begin{bmatrix}
E_y \\[2pt] B_z
\end{bmatrix}_{x+l_i}
=
M(k_i,l_i)
\begin{bmatrix}
E_y \\[2pt] B_z
\end{bmatrix}_{x},
\end{equation}
where the transfer matrix for a uniform layer is
\begin{equation}
M(k_i,l_i) =
\begin{bmatrix}
\cos(k_i l_i) & \dfrac{\omega}{k_i c}\,\sin(k_i l_i) \\[6pt]
-\,\dfrac{k_i c}{\omega}\,\sin(k_i l_i) & \cos(k_i l_i)
\end{bmatrix},
\end{equation}
in CGS units.

To find how the fields evolve across an entire cell of thickness $d$, we note that the fields are continuous across the boundary separating the two layers. As a result, the transfer matrix that maps the field from the left side of the cell to the right side is the product
\begin{equation}
M_{\rm cell} = M(k_2,l_2)\,M(k_1,l_1), \qquad d = l_1 + l_2.
\end{equation}
A straightforward multiplication gives the trace
\begin{eqnarray}
\mathrm{Tr}\,M_{\rm cell}
&=& 2\cos(k_1 l_1)\cos(k_2 l_2) \nonumber \\
&& -\left(\frac{k_1}{k_2}+\frac{k_2}{k_1}\right)\sin(k_1 l_1)\sin(k_2 l_2). \label{trace}
\end{eqnarray}
We look for a solution that satisfies the condition
\begin{equation}
\begin{bmatrix} E_y \\ B_z \end{bmatrix}_{x+d}
= \mu \begin{bmatrix} E_y \\ B_z \end{bmatrix}_{x},
\qquad \mu = e^{i\kappa d}.
\end{equation}
Here, $|\mu|$ directly quantifies the change in wave amplitude across a single cell, which is the same quantity used in the main text to estimate the amplitude reduction per cell for a given density modulation. Since $\det M_{\rm cell}=1$, $\mu$ obeys
\begin{equation}
\mu^2 - (\mathrm{Tr}\,M_{\rm cell})\,\mu + 1 = 0. \label{eq: mu det M}
\end{equation}
Propagating solutions occur when $|\mathrm{Tr}\,M_{\rm cell}|\le 2$, which yields a real $\kappa$. In contrast, $|\mathrm{Tr}\,M_{\rm cell}|>2$ makes $\kappa$ complex, so the field decays exponentially with $x$ — the signature of Bragg reflection.

Even a weak modulation produces reflection. To show this, we consider $n_i = n_0 + \delta n_i$ and use the dispersion relation given by Eq.~(\ref{eq: normal}) where we replace $n_0$ by $n_i$ for layer $i$. The number of particles is conserved in a given cell when the two waves create spikes, so $\delta n_1 l_1 + \delta n_2 l_2 = 0$. Using this relation after expanding the expression for $k_i$ while treating $|\delta n_i|$ as a small parameter, we obtain that $k_1 l_1 + k_2 l_2 \approx \pi$. We use this in Eq.~(\ref{trace}) to find that $|\mathrm{Tr}\,M_{\rm cell}|>2$ for $k_1 \neq k_2$, which corresponds to a complex $\kappa$ and thus an evanescent wave. Therefore, even relatively small spikes cause wave reflection.


\section{Relativistic pair-plasma reflection front} \label{sec:app:rel hole boring}

We follow a standard momentum-balance analysis \cite{Robinson.PPCF.2009} to derive a solution for a relativistic pair-plasma reflection front. The key assumption is that the reflection front is maintained by a circularly polarized electromagnetic wave of normalized amplitude $a_0$. This self-consistent boundary sweeps up all of the pair plasma while moving with a constant velocity $v_b$.

It is convenient to consider the front in a frame moving with a normalized velocity $\beta_b = v_b/c$. We denote the quantities in this frame using the prime symbol ($'$). In this frame, the electromagnetic momentum flux across the plasma boundary is conserved. The flux to the left of the front, assuming complete reflection, is $-E'^2/2\pi$, where $E'=(a_0 m_e c/|e|) \omega'$ and $\omega'=\omega\sqrt{(1-\beta_b)/(1+\beta_b)}$. The upstream particles move toward the front with speed $\beta_b c$. For each species, the density in this frame is $n'=\gamma_b n_0$. Since the particles are reflected, each species contributes twice its incoming flux, and with both electrons and positrons present, the total plasma momentum flux is $-4 n' \gamma_b m_e c^2 \beta_b^2$. Equating the two fluxes gives
\begin{equation}
2 a_0^2 n_{cr} \frac{1-\beta_b}{1+\beta_b}
 = 4 n_0 \gamma_b^{2} \beta_b^2,
\label{eq:pair_front_balance}
\end{equation}
where
\begin{equation}
    \gamma_b=\frac{1}{\sqrt{1-\beta_b^2}}.
\end{equation}

Equation \ref{eq:pair_front_balance} can be used to determine the velocity of the reflection front for given $a_0$ and $n_0/n_{cr}$. Introducing the parameter
\begin{equation}
    \psi \equiv 2 n_0/a_0^2 n_{cr},
\end{equation}
the balance equation reduces to
\begin{equation}
    (1 - \beta_b)^2 = \psi \beta_b^2,
\end{equation}
which shows that $\psi$ is the only parameter controlling the solution. We focus on the regime where $2 n_0/n_{cr} < a_0$, i.e. $\psi < 1$ for $a_0 > 1$. Choosing the positive root (the other corresponds to a negative velocity and is unphysical), we obtain 
\begin{equation}
    \beta_b = \frac{1}{1+\sqrt{\psi}}. \label{eq:beta b}
\end{equation}

\end{document}